\documentclass[prb,twocolumn,showpacs]{revtex4}

\usepackage{graphicx}

\begin{document}

\title{Nonlinear $c$-axis transport in Bi$_2$Sr$_2$CaCu$_2$O$_{8+\delta}$ from two-barrier tunneling}

\author{M. Giura}
\affiliation{Dipartimento di Fisica and Unit\`{a} CNISM,\\
Universit\`{a}  ``La Sapienza'', P.le Aldo Moro 2, 00185 Roma, Italy}

\author{N. Pompeo}
\affiliation{Dipartimento di Fisica ``E. Amaldi" and Unit\`{a} CNISM,\\
Universit\`{a} Roma Tre, Via della Vasca Navale 84, 00146 Roma, Italy}

\author{E. Silva$^*$}
\affiliation{Dipartimento di Fisica ``E. Amaldi" and Unit\`{a} CNISM,\\
Universit\`{a} Roma Tre, Via della Vasca Navale 84, 00146 Roma, Italy}

\date{\today}

\begin{abstract}

Motivated by  the peculiar features observed through intrinsic tunneling spectroscopy of  Bi$_2$Sr$_2$CaCu$_2$O$_{8+\delta}$ mesas in the normal state, we have extended the normal state two-barrier model for the $c$-axis transport [M. Giura et al., Phys. Rev. B  {\bf 68}, 134505 (2003)] to the analysis of $dI/dV$ curves. We have found that the purely normal-state model reproduces all the following experimental features: (a) the parabolic $V$-dependence of $dI/dV$ in the high-$T$ region (above the conventional pseudogap temperature), (b) the emergence and the nearly voltage-independent position of the ``humps" from this parabolic behavior lowering the temperature, and (c) the crossing of the absolute $dI/dV$ curves at a characteristic voltage $V^\times$. Our findings indicate that conventional tunneling can be at the origin of most of the uncommon features of the $c$ axis transport in  Bi$_2$Sr$_2$CaCu$_2$O$_{8+\delta}$. We have compared our calculations to experimental data taken in severely underdoped and slightly underdoped  Bi$_2$Sr$_2$CaCu$_2$O$_{8+\delta}$ small mesas. We have found good agreement between the data and the calculations, without any shift of the calculated $dI/dV$ on the vertical scale. In particular, in the normal state (above $T^\ast$) simple tunneling reproduces the experimental $dI/dV$ quantitatively. Below $T^\ast$ quantitative discrepancies are limited to a simple rescaling of the voltage in the theoretical curves by a factor $\sim$2. The need for such modifications remains an open question, that might be connected to a change of the charge of a fraction of the carriers across the pseudogap opening.

\end{abstract}

\maketitle

\section{Introduction}
\label{intro}
Tunneling is an inherent phenomenon in highly anisotropic cuprate superconductors.\cite{kleinerPRB94} Due to their layered structure of conducting sheets intercalated by insulating blocks, superconductivity in these compounds is often described in terms of thin superconducting layers coupled via Josephson tunneling. Similarly, tunneling between the layers is expected to affect the transverse ($c$-axis) conductivity in the normal state, in competition with thermally activated interlayer hopping. If the interlayer barriers are not dramatically high, the latter process is a natural consequence at sufficiently high temperatures.\\
In this sense, measurements of the $c$-axis conductivity add to the various other probes of the transverse quasiparticle transport: intrinsic tunneling spectroscopy, scanning tunneling spectroscopy, angle-resolved photo emission spectroscopy (ARPES). In particular, when an intrinsic tunnel junction exists, $\rho_c$ is a good measure of the density of states (DOS) around the Fermi level (low energy, in the spectroscopic language).\\
Experimental investigations of the transverse properties, such as the differential conductivity $dI/dV$ and ARPES (which probes the $c$-axis properties at selected momentum), reveal the existence of different energy scales and their evolution with doping and temperature. The most evident finding is the appearance of sharp features at the superconducting transition (``peaks") at typical energies $\Delta_{S}$. The temperature evolution of $\Delta_{S}$ which, starting from zero, increases upon decreasing $T$ from $T_c$, ensures the identification with the superconducting gap.\\
Much less obvious is the appearance of weaker features (``humps") at larger energies $\Delta_{pg}$, which is often identified with the pseudogap. In this case the temperature dependence  is much less pronounced. Sometimes a constant $\Delta_{pg}(T)$ is observed decreasing $T$, while only the number of carriers is seen to decrease. The interpretation of the connection between $\Delta_{S}(T)$ and $\Delta_{pg}(T)$ is still a very debated subject\cite{stajicPRB03} and a crucial issue for the understanding of high-temperature superconductivity.\\
Measurements of the differential conductivity $dI/dV$ gave a substantial contribution in bringing the mentioned features into a clearer light, at least on experimental grounds. The superconducting gap has been identified with the voltage $V_p$ where the peaks in $dI/dV$ appear.\cite{krasnovPRL00,krasnovPRB02} A very weak temperature dependence of the bias voltage where the humps appeared, $V_h$, has been consistently reported in several papers.\cite{suzukiPRL99,krasnovPRL00, suzukiPRL00,krasnovPRB02} In addition, it was shown that $V_h$ was insensitive to a magnetic field, while $V_p$ was reduced to zero by large magnetic fields,\cite{krasnovPRL01} suggesting a different nature of the phenomena involved.\\
For what concerns the description of the data, the usual procedure has been to include some specific shape of the DOS in the model to fit the shape of the $dI/dV$ curves. In particular, below $T_c$ it is now customary to use some BCS-like expression of the superconducting gap to obtain estimates of the maximum gap and of the scattering rate from fitting of the $dI/dV$ curves.\cite{zasadzinskiPRL01} This approach has been extended to the fit of the humps. Taking two different gaps the $dI/dV$ at 4 K were reproduced.\cite{romanoPRB06} A continuation of the fitting of the $dI/dV$ above $T_c$ with a nearly $T-$independent gap has been also reported.\cite{zhaoCM07} The $T-$independence of $V_h$ (Ref.\onlinecite{krasnovPRL00}) and, consequently, of the pseudogap if one identifies the experimental feature with a gap in the DOS, would imply that $\Delta_{pg}$ is an energy scale for a crossover instead that for a true transition.\\
Proceeding toward the normal state, we note that  another interesting feature appears. The $dI/dV$ curves at high $T$ (and sufficiently large doping) present a very anomalous downward parabolic behavior,\cite{krasnovPRB02,suzukiPRL99,ozyuzer00} which has been treated up to now as an unspecified ``background".\\
In this paper we address the issue of the normal state of layered superconductors, focussing on Bi$_2$Sr$_2$CaCu$_2$O$_{8+\delta}$ (BSCCO), where most of the experiments have been performed. We build up a model for the $c$-axis $dI/dV$ based on the presence of two different energy barriers, where each one gives rise to tunnel and thermally activated currents.\cite{giuraPRB03, giuraPRB04, giuraSUST07} We show that this simple model, based solely on the existence of two energy barriers and without any particular features in the DOS, can reproduce many aspects of the experimental $dI/dV$. In particular, the parabolic behavior is reproduced quantitatively. The presence (in overdoped BSCCO) or absence (in underdoped BSCCO) of such high-$T$ parabolic $dI/dV$ finds an explanation in the simulations with the different balance between tunneling and thermal activation, depending on the height of the energy barriers. Moreover, the balance between thermal and tunnel currents across the barriers gives rise to humps in the $dI/dV$, suggesting that the humps arise from purely normal-state tunneling phenomena. The nearly constant position of the humps as a function of the temperature in experimental data, as well as the crossing of the $dI/dV$ at a characteristic voltage in underdoped samples, are reproduced by the simulations. Finally, a simple rescaling of the voltage scale by a factor $\sim$2 is sufficient to bring the simulations in accurate quantitative agreement with experimental data taken in BSCCO mesas.\\
The paper is organized as follows.  In Section \ref{model} we recall the two-barrier model for  the
$c$-axis conduction, focussing on the extension of the model to the description of nonlinear $dI/dV$ vs. $V$ curves. We find that several experimental features are recovered by the simple model. In Section \ref{disc} we discuss data taken in BSCCO mesas in terms of the two-barrier model. A short summary is presented in Section \ref{conc}.\\
\section{$dI/dV$ in the two-barrier model}
\label{model}
\begin{figure}
     \includegraphics[width=5cm]{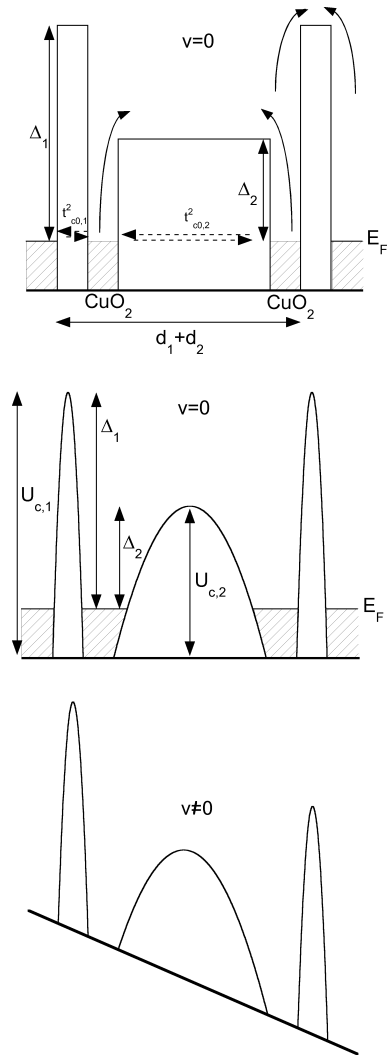}
\caption{Upper panel: sketch of the energy barrier profile used to describe the linear (small voltage) $\rho_c$. The abscissa is the distance along the crystallographic $c$ axis. Middle panel: modified energy profile chosen for the nonlinear transport. Lower panel: effect of an applied electric field (voltage) on the barriers. Note that the two barriers undergo different changes (uneven voltage drops).}
\label{fig_barriers}
\end{figure}

It is generally believed that the $c$-axis conduction in BSCCO is determined by the layered structure. In particular, tunneling certainly plays a major role. To address this issue we have introduced and developed \cite{giuraPRB03,giuraPRB04,giuraSUST07} a phenomenological model for the out-of-plane resistivity $\rho_{c}$ in layered superconductors.  In particular, we have proposed that the two-layer structure of BSCCO could give rise to two barriers for the $c$ axis conductivity. The overall conductivity would then be the composition of thermal and tunnel transport across each of the barriers. A sketch of the barriers considered is presented in Figure \ref{fig_barriers}. Clearly, this model directly applies to the true normal state, i.e. the state where no other energy scales appear (due to, e.g., the opening of the pseudogap or of the superconducting gap\cite{notaTc}). More specifically, no particular \textit{ad hoc} shape of the density of states is considered. Regarding the scattering process involved, we have considered that tunneling across the barriers could be reduced due to incoherent in-plane scattering, along the lines of Ref.\onlinecite{kumarPRB92}. This process reduces the average interlayer tunneling hopping rate, thus leading to an increased charge localization in the $(a,b)$ planes. In particular, in-plane inelastic phononic scattering given by the scattering time $\tau$ reduces the tunneling matrix element $t_c$ to $(2\tau/\hbar)t_{c}^{2}\ll t_{c}$. In the linear response (vanishing voltage), the resistivity in the two-barrier model is then given by the simple expression:\cite{giuraPRB03,giuraSUST07}\\
\begin{eqnarray}
\rho_{c,n}& = & \frac{1}{d_{1}+d_{2}}\sum_{i=1,2} \left[ \frac{t_{c0,i}^2}
{a\rho_{ab,n}+b}+\beta e^{-\Delta_i/k_BT} \right]^{-1}
\label{rholin}
\end{eqnarray}
where the first and second term in square brackets represent tunnel and thermal activation over the $i$th barrier, respectively. Thus, the model represents a series of two elements, each of them is a parallel of two channels. Scattering processes enter through $\tau^{-1}\propto a\rho_{ab,n}$ (in-plane scattering) and $b$ (out-of-plane scattering), where $\rho_{ab,n}$ is the $(a,b)$ plane resistivity in the normal state, $d_1 =$~3 \AA~ and $d_2 =$~12 \AA~ are the spacings between the CuO$_2$ layers, and the height $\Delta_i$ of the $i$-th barrier enters through $t_{c0,i}^2\propto\exp[2d_i\sqrt{2m^{*}\Delta_i}/\hbar]$ (with $m^{*}=4.6 m_{e}$, Ref.\onlinecite{poole}). For small applied voltages  one can take a rectangular shape for the energy barriers without loss of generality, as depicted in Figure \ref{fig_barriers}, upper panel. With this simple model we were able to reproduce the doping dependence of $\rho_c(T)$ in BSCCO quantitatively, and in particular the evolution from a nearly linear $\rho_c(T)$ in overdoped samples to the ``semiconducting" behavior in underdoped samples. In particular, the model was able to reproduce the local minimum of $\rho_c(T)$ which appears at optimal doping and slight underdoping.\cite{giuraPRB03} The details, including the crossover from thermal to tunnel transport by varying the temperature and the height of the barrier, have been worked out and discussed previously.\cite{giuraPRB03,giuraSUST07} In the following we focus on the extension of this model to nonlinear transport.\\
The problem of nonlinear tunneling is a long standing issue in solid state physics,\cite{simmonsJAP63, simmonsJAP64a, simmonsJAP64b, chowJAP65,wattamaniukPRL75,mahan} and it is still far from being fully solved.\cite{mahan} Due to the extreme complexity of the problem, we use an approach as simple as possible to incorporate the effect of finite voltages into our model. In particular, we have to incorporate (a) the deformation of the energy barriers due to increasing voltage, (b) the existence of two barriers ``in series", which implies nonequal, $V$-dependent voltage drops on each barrier, leaving unaffected (c) the role of the in-plane scattering, which is a physical foundation of the linear model employed up to now.\\
In a nonlinear problem, it is expected that the applied voltage affects the shape and height of the energy barrier.\cite{ma} This effect cannot be taken into account by rectangular barriers. Among the many possible choices, we have found that parabolic shapes for the energy barriers led to an analytical, albeit complicated, expression for the tunneling current. We checked that the results did not change appreciably with other barrier shapes, such as trapezoidal or quartic or higher order polynomials. While the simulations were not very sensitive to the detailed shape of the barriers, two requirements emerged from extensive simulations: first, the barriers needed to vary their height with the applied electric field, that is rectangular barriers were found to be unsuitable for an even qualitative reproduction of the experiments. Second, we found that the existence of two different barriers ensured a nearly quantitative reproduction of the experimental  $dI/dV$. Thus, we chose the energy profile sketched in Figure \ref{fig_barriers} (middle panel) to represent the two-barrier nonlinear model, with: $U_{i}(x)=U_{ci}-\frac{\alpha}{2}(x-\frac{d_{i}}{2})^2$ in the interval $0<x<d_{i}$, and zero otherwise.  With respect to the previously defined $\Delta_i$, one has $U_{c,i} = \Delta_i+E_F$ and $\alpha=8U_{ci}/d_{i}^{2}$. It is important to stress that the choice of parabolic barriers does not introduce additional parameters with respect to the rectangular profile.\\
The effect of an applied electric field $\mathcal{E}$ is exemplified in Figure \ref{fig_barriers}, lower panel: the barriers are lowered and distorted, and eventually undergo a breakdown (regime which is out of the scope of the present paper). The energy profile changes to: $U_{i}^{\prime}(x)=U_{i}(x)-e\mathcal{E}_i x$, where $\mathcal{E}_i=V_i/d_i$ is the electric field within the $i-$th barrier, where the voltage drops by $V_i$.\\
The voltage induced change of shape determines significant changes in both thermal and tunnel currents. The thermal current density across the $i-$th barrier can be written as:\cite{ma}
\begin{equation}
    J_{th,i}(V_i)=-\frac{eDn\left(1-e^{eV_{i}/k_{B}T}\right)}
    {\int_{0}^{d_{i}}e^{\left[U_{i}^{\prime}(x)-E_{F}\right]/k_{B}T}dx}
\label{Jth}
\end{equation}
where $D$ is the diffusion coefficient, and $n=const$ is the volume density of carriers outside the barrier. This expression can be solved analytically\cite{giuraSUST07} to yield a closed form for $J_{th,i}$.\\
To write down the expression of the tunnel current one needs the transparency $T_{ci}^2$ of the $i-$th barrier. We incorporate the effect of the electric field in analogy to the field effect. The final expression depends on the overall energy, and is:\cite{giuraSUST07}
\begin{eqnarray}
T_{ci}^{2}(E) =
	\exp
	\left\{
	  -\frac{\pi\sqrt{2m^{*}U_{c,i}}d_{i}}{4\hbar}
	  	\left[
     		\left(1-\frac{eV_{i}}{4U_{c,i}}\right)^{2}-\frac{E}{U_{c,i}}
    		 \right]
	\right\}
\label{tcnonlinear}
\end{eqnarray}
for $E<U_{ci}\left(1-\frac{eV_{i}}{4U_{ci}}\right)^{2}$ and $T_{ci}^2=1$ otherwise. Assuming that the in-plane scattering affects the tunneling matrix\cite{kumarPRB92} and the transparency similarly, and that it holds in the nonlinear regime, the tunnel current is written as follows:
\begin{equation}
    J_{tun,i}(V_{i})=-
    \int_{0}^{\Delta_{i}+E_{F}}
    \frac {T_{ci}^{2}(E)}{e(a\rho_{ab}+b)}
    \left[
    f(E)-f(E+eV_{i})
    \right] dE
\label{Jtun}
\end{equation}
where $f$ is the Fermi function.\\
\begin{figure}
\includegraphics[width=6cm]{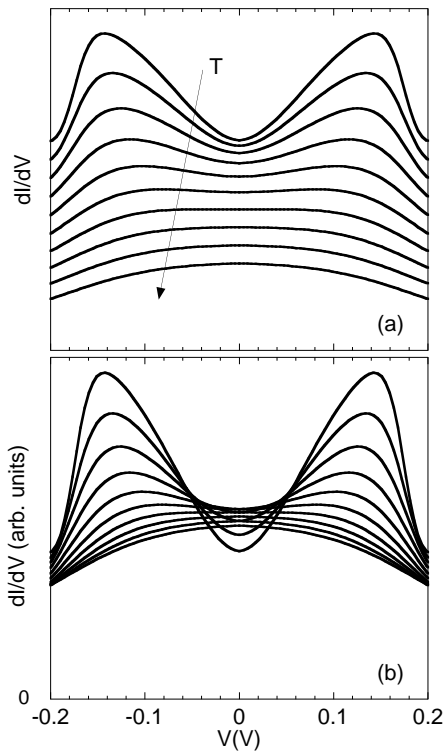}
\caption{Simulations of the $dI/dV$ curves by means of the two-barrier model, using parameters typical for an overdoped BSCCO. Curves are calculated every 25 K for $T\leq$ 300 K. With reference to Figure \ref{fig_barriers}, we used $\Delta_1=$ 2710 K, $\Delta_2=$ 610 K.
Panel (a): calculated curves arbitrarily shifted vertically. Panel (b): same curves, on absolute scales. Note the clear parabolic behavior and the emergence of the humps with lowering temperature. In real samples the crossing would be masked by the superconducting transition.}
\label{fig_simulover}
\end{figure}
To complete the model we need to stress that experiments are usually performed applying an external potential $V$ to the sample (typically a mesa crystal, composed of $N$ cells). It is true that the total voltage distributes evenly across the $N$ cells, so that it is customary to report the measurements as a function of the voltage per cell, $v=V/N$. However, within our model each cell is composed of two unequal barriers, so that  $v$ is the sum of the potential drops over the two barriers: $v=V_{1}+V_{2}$, with $V_{1}/V_{2}$ variable with temperature or with $v$ itself. The calculations must then be performed selfconsistently, with the requirement of the same total current through each barrier: $J=J_{1}=J_{2}$, with $J_{i}=J_{th,i}+J_{tun,i}$ (the cross section $S$ is the same throughout the sample).\\
Before presenting some of the results of the model, we recall that the model itself contains  five adjustable parameters: $a, b, \beta, \Delta_{1}, \Delta_{2}$, being $\rho_{ab,n}$ a measured quantity. None of the parameters are temperature dependent, while all can be doping dependent. Moreover, several constraints act to further reduce the freedom of choice of the parameters. First, $a, b, \beta$ can be combined so that only two of them affect the shape of the calculated curves, the third acting as a mere scale factor. Second, the fits of $\rho_{c}$ at low voltage set a lower bound for $\Delta_{1}-\Delta_{2}$. Once the set of parameters is fixed, one directly generates the full $T$-dependent family of curves $dI/dV$ vs. $V$, and obviously $\rho_c(T)$.\\
We would like to stress that the model sketched up to now is a pure ``normal state" model: it takes into account thermal and tunnel transport across a series of two barriers between an otherwise conventional conductor, with the only specific introduction of the in-plane scattering effect on the transparency. In particular, this model does not introduce any specifically tailored DOS in the tunnel current, Eq. (\ref{Jtun}), with implicit or explicit introduction of a gap opening at a certain temperature, and contains only the energies given by $\Delta_1$, $\Delta_2$ and $E_F$. Moreover, the model does not concern the in-plane transport. Instead, it takes $\rho_{ab}$ from the experiments, assuming that it reflects the in-plane scattering (through e.g., the ratio of the parameters $b/a$).\\
We now illustrate the predictions for the nonlinear $dI/dV$ curves, presenting some simulations with sets of parameters typical for overdoped samples and for underdoped samples. Following the common habit of presentation of the data, first we will present our simulations for $dI/dV$ translating them arbitrarily on the vertical scale. By replotting the same curves on an absolute scale we will show that significant information is lost in the most common representation.\\
In Figure \ref{fig_simulover} we report the voltage dependence of the differential conductivity with parameters chosen to describe $\rho_c$ in an overdoped sample. In Figure \ref{fig_simulover}a we plot the data as it is customary, that is with some arbitrary vertical translation, while in  Figure \ref{fig_simulover}b the same curves are replotted in an absolute scale. We first notice that the present model reproduces an otherwise unexplained feature of the experiments: the parabolic behavior of $dI/dV$ in the normal state. In fact, this behavior is routinely observed in the experiments,\cite{krasnovPRB02,suzukiPRL99,ozyuzer00} but it is often ignored. Sometimes the data are normalized to this ``background",\cite{zasadzinskiPRL01} attributing it to some instrumental feature or to self-heating (however, we note that the parabolic shape is observed also when self-heating is avoided by ns time-scale short-pulse measurements\cite{anagawaPRB06}). Clearly, such a normalization possibly removes some significant information from the data.\\
A second striking result is the emergence of the well-known ``humps" from the parabolic background. This is a very important result of our simulations: without any additional energy scale (such as a gap introduced in the DOS) we reproduce such humps only by virtue of the balance between the tunnel and thermal current across the barriers. Moreover, the simulations reproduce the nearly $T$-independence of the position of the hump, $V_h$, observed in many overdoped samples, and the increase of the hump height with decreasing temperature.\cite{krasnovPRB02,krasnovPRL00,anagawaPRB06}\\
\begin{figure}
\includegraphics[width=6cm]{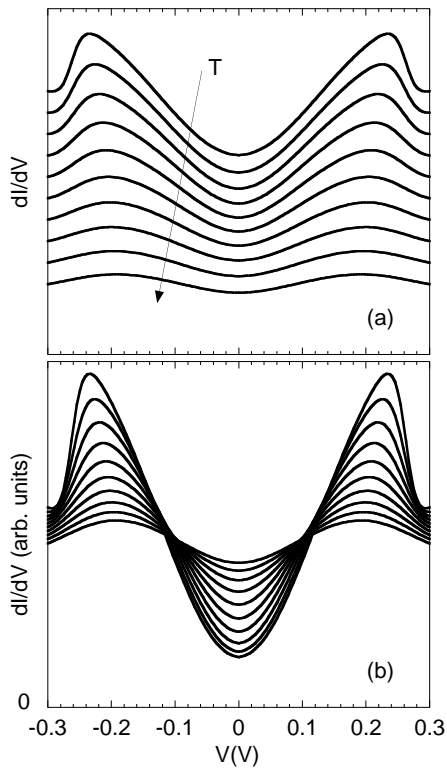}
\caption{As in Figure \ref{fig_simulover}, for an underdoped sample. Curves are calculated every 25 K for $T\leq$ 300 K. Barriers heights are $\Delta_1=$ 3270 K, $\Delta_2=$ 1170 K.
Panel (a): calculated curves arbitrarily shifted vertically. Panel (b): same curves, on absolute scales. Note the crossing of all the curves at $V^\times$.}
\label{fig_simulunder}
\end{figure}
Changing parameters to those typical for underdoped samples several other features are revealed. In Figure \ref{fig_simulunder} we report the simulations corresponding to underdoping. Again, several familiar experimental features emerge from the model. The humps still rise from the background, with little temperature dependence of $V_h$, as in the previous simulations. Of great interest is the plot of the curves on absolute scale: we see that all the simulated curves cross each other at a typical voltage $V^{\times}$. This feature is appreciable only in the absolute scale. It is noteworthy that this is exactly the experimental behavior exhibited by underdoped mesas, as reported by many groups.\cite{krasnovPRB02,yamadaPRB03}\\
Finally, note that in Fig.s \ref{fig_simulover},\ref{fig_simulunder} the scales are comparable to those in typical experiments, thus suggesting that quantitative fitting is possible.\\
The results of the simulations obviously call for precise measurements and direct comparison with the experiments, task accomplished in the following Sections.
\section{Results and Discussion}
\label{disc}
\begin{figure}
\includegraphics[width=6cm]{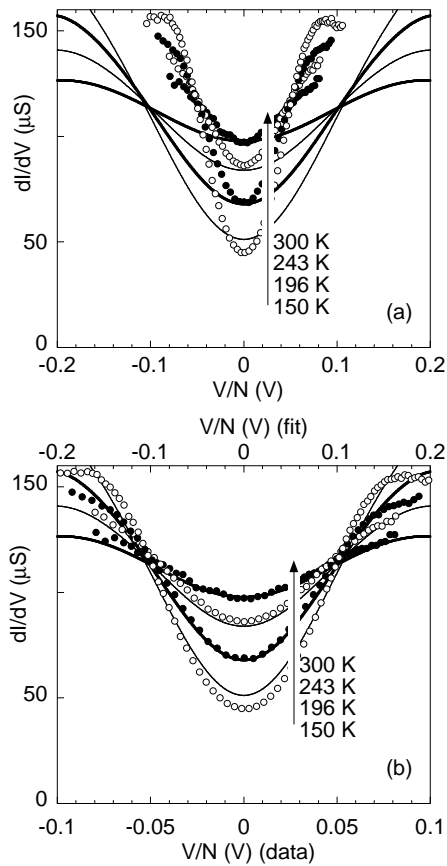}
\caption{Comparison of the model to the experimental data in an underdoped sample ($\delta\leq$0.2135). Panel (a): data and fits are reported on absolute scales. Barriers heights as in Figure \ref{fig_simulunder}. Panel (b): the voltage scale for the fits has been divided by $\gamma=$2. By the simple rescaling of the voltage scale the fits reproduce the data well. The qualitative features are all contained in the model.}
\label{fig_under}
\end{figure}

\begin{figure}
\includegraphics[width=6cm]{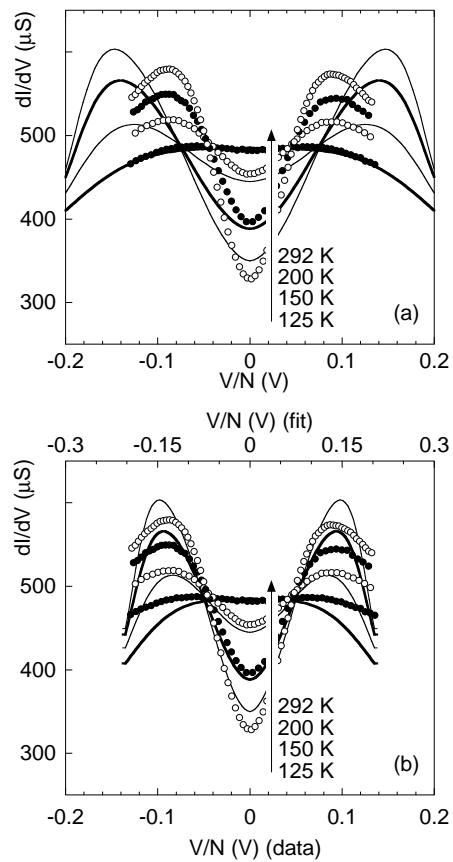}
\caption{Comparison of the model to the experimental data in a slightly underdoped sample ($\delta=$0.24). Panel (a): data and fits are reported on absolute scales. Note that the parabola at high $T$ is reproduced. The barriers heights are: $\Delta_1=$ 2880 K, $\Delta_2=$ 780 K.
Panel (b): the voltage scale for the fits has been divided by $\gamma=$2. Rescaling of the voltage brings the fits in agreement with the data below the temperature where the humps appear, but the fit of the parabolic behavior is lost.}
\label{fig_slightly}
\end{figure}

In this Section we compare our simulations with experimental data taken in small BSCCO mesas at Stockholm University.\cite{krasnovdata} Experimental details for data taken on similar mesas can be found in Ref.\onlinecite{krasnovARXIV08}.\\
One of the key points is the determination of the barrier heights and of the other parameters needed to reproduce both the linear (small voltage) $\rho_c(T)$ and the nonlinear differential conductance $dI/dV$ curves. The parameters are determined as follows. Using the procedure illustrated previously,\cite{giuraSUST07} we fitted $\rho_c(T)$ in a large number of BSCCO single crystals at different doping $\delta$. The data were taken with an eight-terminal configuration\cite{espositoJAP00} or from the literature.\cite{watanabePRL97} The resulting parameters exhibited very regular doping dependences, such as a linearly decreasing $\Delta_2(\delta)$.\cite{giuraPRB03} By comparing the data taken in the mesa with the full body of the preexisting elaborations we identified the parameters to be used. The $dI/dV$ curves were then calculated directly, without further adjustments. We remark again that we will compare the simulation and the data on an absolute scale, without any vertical translations.\\
On general grounds, we anticipate that the two-barrier model reproduces the linear and nonlinear behavior in the (supposedly) true normal state with great accuracy, while some quantitative discrepancy appear crossing the temperature $T^*$ where the pseudogap is supposed to be located.\\
To investigate the pseudogap region we use the data taken on a severely underdoped sample with $\delta\leq$0.2135. Figure \ref{fig_under}a reports the comparison with the model. The calculations are performed using parameters appropriate to that doping. One can see that the shape of the calculated curves is very close to the data, while the voltage scale compared to the experimental data appears stretched. Interestingly, rescaling the voltage of the calculated curves by a factor $\gamma =$ 2 the agreement becomes impressive, as shown in Figure \ref{fig_under}b. We note that the qualitative features do not depend on the rescaling: the model reproduces the emergence and the evolution of the humps and, what it is particularly relevant, the crossing of the curves at a typical voltage. The voltage rescaling brings in quantitative agreement the model and the data. To appreciate the result we recall again the absolute vertical scale.\\
The need for a voltage rescaling is a feature to be investigated. The best approach is to compare the model to data in the slightly underdoped sample, where all the characteristic features are present: parabolic characteristics at high $T$, emergence of the humps, crossing of the curves in the ``hump" region. Figure \ref{fig_slightly} reports the comparison of the model with the data in a sample with $\delta=$0.24. As it can be seen in Figure \ref{fig_slightly}a the parabolic behavior is correctly reproduced, again without any voltage rescaling. However, when the humps emerge and the crossing voltage appears, the model describes (nearly quantitatively) the data only after a voltage rescaling by a factor $\gamma \approx$ 1.5, similar to the factor needed in the severely underdoped sample. Noteworthy, the appearance of the scaling factor is connected to the apperance of the humps (and of the crossing voltage). We stress again that no modification of the DOS is included in the model: the evolution from parabola to humps appears only as a consequence of the balance between thermal and tunneling transport, and of the dynamically adjusting voltage drop between the two barriers. Moreover, the full $dI/dV$ are fitted, \textit{in absolute voltage and conductance scale}. This result is even more remarkable since only $T$-independent fitting parameters are involved.\\
It can be concluded that several typical features of the data (in particular, the parabolic $dI/dV$ and the emergence of the humps) are reproduced by the two-barrier calculations. When the humps appear, we find that the calculated $dI/dV$s require a scaling of the voltage by a factor $\gamma >$1 to match quantitatively the data. By further lowering the temperature, the rescaled simulations track the \textit{position} of the experimental features (humps, crossing), even if the experimental $dI/dV$s show a reduced dynamics with respect to the simulation (as expected in a pseudogap state). We believe that, while the humps are a feature inherent in the layered structure of BSCCO and, as such, captured by the simple two-barrier model, the additional physics that develops in the pseudogap state needs a specific incorporation into the model. The role of the voltage scaling factor is still unclear, but it is evidently connected to the electronic transformation at $T^*$. To the best of our efforts, the rescaling cannot be achieved changing the other parameters in the tunnel and thermal current. We mention that, on purely numerical grounds, a rescaling of the electric charge $e\rightarrow\gamma e$ produces the desired rescaling of the curves. However, before drawing strong conclusions on the charge of the carriers much more work would be needed. As an example, it should be considered that any ``preformed pair" scenario would certainly affect the DOS and ultimately change the shape of the $dI/dV$s. We shall leave further work to future studies, but we emphasize again how a simple, purely ``normal-state" model as the one here depicted can reproduce many features of the experiments with surprising success.
\section{Summary}
\label{conc}
Summarizing, we have extended the two-barrier analysis of the $c$-axis transport for the normal state of BSCCO to the $dI/dV$ curves, taken in mesas at different doping level. We have found that many of the peculiar features of such curves are reproduced by the two-barrier model, namely: (a) the parabolic $dI/dV$ in the high-$T$ region (above $T^\ast$), (b) the emergence of the ``humps" from this parabolic behavior and their nearly constant position (on the voltage scale) lowering the temperature, and (c) the crossing of the \textit{absolute} $dI/dV$ curves at a characteristic voltage $V^\times$. Our findings indicate that ``conventional" transport across the energy barriers (tunneling and thermal activation) could be at the origin of most of the uncommon features of the $c$-axis transport in BSCCO. In the normal state (above $T^\ast$) simple tunneling and thermal activation reproduce the experimental $dI/dV$ quantitatively. Below $T^\ast$ quantitative discrepancies appear. A rescaling of the charge of the carriers and a reduction of the dynamic range bring the calculations in quantitative agreement with the experiments. The clarification of the normal state in terms of single electron tunneling and thermal activation could help to identify the possible scenarios for the physics of the pseudogap state. The need for such modifications remains an open question that we intend to tackle in a future work.
\begin{acknowledgments}
We gratefully thank V. Krasnov at Stockholm University for sending us unpublished data on BSCCO mesas and for stimulating discussions.
\end{acknowledgments}


\begin{thebibliography}{99}
\bibitem{notaTc}
While the extension of this model in the superconducting state is beyond the scope of the present work, we note that below $T_c$ the two-barrier model would show a series of two SIS junctions, most probably with the thin barrier (of height $\Delta_1$ in the notation of the main text) exhibiting a much higher critical current.
\end{thebibliography}
\end{document}